\documentclass[twocolumn,showkeys,nopacs,prl,floatfix]{revtex4}
\usepackage{epsfig}

\begin{document}

\title{Fingerprinting Hysteresis}

\author{Helmut G.~Katzgraber$^a$, Gary Friedman$^b$, and G.~T.~Zim\'anyi$^c$}
\affiliation{$^a$Theoretische Physik, ETH H\"onggerberg, 
CH-8093 Z\"urich, Switzerland}
\email{katzgraber@phys.ethz.ch}
\altaffiliation{Fax: +41 1 633 11 15 (author responsible for further
correspondence)}

\affiliation{$^b$Drexel University, ECE Department, Magnetic Microsystems
Laboratory, Philadelphia, PA 19104}

\affiliation{$^c$Department of Physics, University of California,
Davis, California 95616}

\date{\today}

\begin{abstract}
We test the predictive power of first-oder reversal
curve (FORC) diagrams using simulations of random
magnets. In particular, we compute a histogram of the
switching fields of the underlying microscopic
switching units along the major hysteresis loop, and
compare to the corresponding FORC diagram.  We find
qualitative agreement between the switching-field
histogram and the FORC diagram, yet differences are
noticeable. We discuss possible sources for these
differences and present results for frustrated systems
where the discrepancies are more pronounced.
\end{abstract}


\maketitle

\section{Introduction}
\label{introduction}

The conventional
methods \cite{he:92,proksh:94,hedja:94} to characterize
magnetic interactions in hysteretic systems, such as
the $\delta M$ method \cite{che:92,el-hilo:92}, utilize
isothermal remanent magnetization (IRM) and dc
demagnetization remanence (DCD) curves based on the
Wohlfarth relation \cite{wohlfarth:58}. Recently, FORC
diagrams \cite{pike:99,katzgraber:02b} have been
introduced to study hysteretic systems. Their extreme
sensitivity has helped to ``fingerprint'' several
experimental systems as well as theoretical models
ranging from geological samples and recording media to
paradigmatic models of random magnets and spin
glasses \cite{katzgraber:02b}.

In this work we perform numerical simulations of
random magnets (and spin glasses) in order to test the
predictive power of FORC diagrams by comparing to a
histogram of up- and down-switching fields of the
underlying switching units along the major hysteresis
loop.

The aforementioned re-parametrization of the major
hysteresis loop (switching-field histogram) displays
the information carried by the major loop in a more
comprehensive way and provides a good comparison to
the FORC diagram. We find, that the major-loop
behavior predicts the minor-loop behavior captured by
FORC diagrams well. We present a comparison of both
distributions and discuss some differences between
them. We argue that switching-field histograms are
useful to study hysteretic systems in more detail than
with conventional methods due to their simplicity and
ease to compute.

\section{Model \& Algorithm}
\label{model}

The Hamiltonian of the random-field Ising model (RFIM)
is given by \cite{ji:92}
\begin{equation}
{\mathcal H}= \sum_{\langle i,j \rangle} J_{ij}S_iS_j -
                  \sum_i h_i S_i - H  \sum_i S_i \,.
\label{eq:hamilton}
\end{equation}
Here $S_i = \pm 1$ are Ising spins on a square lattice
of size $N = L^3$ in three dimensions with periodic
boundary conditions. The interactions between the
spins are uniform ($J_{ij} = 1$) and nearest-neighbor,
and $H$ represents the externally applied field.
Disorder is introduced into the model by coupling the
spins to site-dependent random fields $h_i$ drawn from
a Gaussian distribution with zero mean and standard
deviation $\sigma_{\rm R}$.

We simulate the zero-temperature dynamics of the RFIM
by changing the external field $H$ in small steps
starting from positive saturation. After each field
step we compute the local field $f_i$ of each spin:
\begin{equation}
f_i=\sum_{j} J_{ij}S_j - H -h_i\; .
\label{eq:local_field}
\end{equation}
A spin is unstable if it points opposite to its local
field, i.e., $f_i \cdot S_i < 0$. We then flip a
randomly chosen unstable spin and update the local
fields at neighboring sites. This procedure is
repeated until all spins are stable.

For the rest of this work we set $L = 50$ ($N =
125000$ spins) and $\sigma_{\rm R} = 5.0$, unless
otherwise specified. The different figures are
calculated by averaging over 5000 disorder
realizations in order to reduce finite-size effects.

\section{FORC Diagrams}
\label{forc}

In oder to calculate an FORC diagram, a family of First
Order Reversal Curves (FORCs) with different reversal
fields $H_{\rm R}$ is measured, with $M(H, H_{\rm R})$
denoting the resulting magnetization as a function of
the applied and reversal fields.  Computing the mixed
second order derivative \cite{dellatorre:99,pike:99}
\begin{equation}
\rho(H, H_{\rm  R})= -\frac{1}{2}
[{\partial}^2 M/{\partial} H {\partial} H_{\rm  R}]
\label{eq:rho}
\end{equation}
and changing variables to $H_{\rm c}=(H-H_{\rm R})/2$
and $H_{\rm b}=(H+H_{\rm R})/2$, the local coercivity
and bias respectively, yields the ``FORC
distribution'' $\rho(H_{\rm b}, H_{\rm c})$. FORC
diagrams resemble the commonly known Preisach
diagrams \cite{preisach:35,mayergoyz:86}, yet they are
model-independent and therefore more general.

\begin{figure}[tbp]
\begin{center}
\includegraphics[scale=0.38]{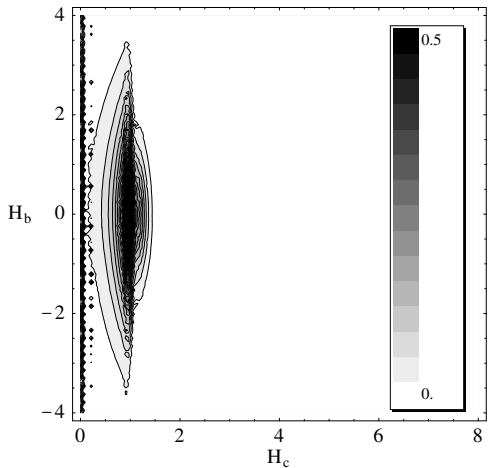}
\end{center}
\vspace*{-0.5cm}
\caption{
FORC diagram of the RFIM for disorder $\sigma_R =
5.0$, well above the critical
disorder \cite{perkovic:99} in three dimensions. Note
the pronounced vertical feature at $H_{\rm c} = 1$
with a wake extending to $H_{\rm c} = 0$ which
corresponds to multi-domain nucleation in the sample.  
The dots along the $H_{\rm b}$-axis are numerical
noise (no data smoothing). 
\label{fig1}
}
\end{figure}

Figure \ref{fig1} shows an FORC diagram of the RFIM at
high disorder strength ($\sigma_{\rm R} = 5.0 >
\sigma_{\rm crit} \approx 2.16$) \cite{perkovic:99}.
Note the vertical ridge at $H_{\rm c} \approx 1$
reminiscent of domain-wall
nucleation \cite{drossel:98}. A vertical cross-section
of the ridge ($H_{\rm c} = 1$)  mirrors the
distribution of the applied random fields, because
these can be viewed as a distribution of random biases
acting on the spins when $\sigma_{\rm R} \gg
\sigma_{\rm crit}$. We have tested this in detail by
selecting the random fields from a box distribution.
The resulting FORC diagram is qualitatively similar to
the Gaussian case, yet a vertical cross-section of the
ridge is box-shaped. This is not evident by studying
the major hysteresis loop for different
disorder-distribution shapes and illustrates the
advantages of the FORC method over conventional
approaches for studying hysteretic systems.

\section{Switching-Field Histograms}
\label{sfh}

In order to test the predictive power of FORC
diagrams, we simulate the RFIM with the zero-temperature
dynamics described in Sec.~\ref{model}
and store the up- and down-switching
fields of the spins along the major hysteresis loop.
We then create a histogram of the number of flipped
spins for a given pair of up- ($H_{\uparrow}$) and
down-switching fields ($H_\downarrow$). By changing
the variables to the coercivity [$H_{\rm c} =
(H_{\uparrow} - H_\downarrow)/2$] and bias [$H_{\rm b}
= (H_{\uparrow} + H_\downarrow)/2$] of each spin, we
obtain a distribution of the coercivities and biases
of the spins in the system along the major hysteresis
loop.

\begin{figure}[tbp]
\begin{center}
\includegraphics[scale=0.38]{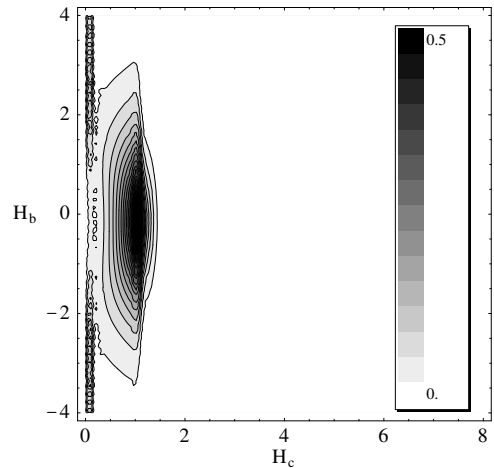}
\end{center}
\vspace*{-0.5cm}
\caption{
Switching-field histogram of the three-dimensional
RFIM with $\sigma_{\rm R} = 5.0$. Note the close
resemblance to the FORC diagram presented in
Fig.~\ref{fig1}. Because no derivatives of the data
are required, the contours are much smoother than in
the case of an FORC diagram.
\label{fig2}
}
\end{figure}

\begin{figure}[tbp]
\begin{center}
\includegraphics[scale=0.38]{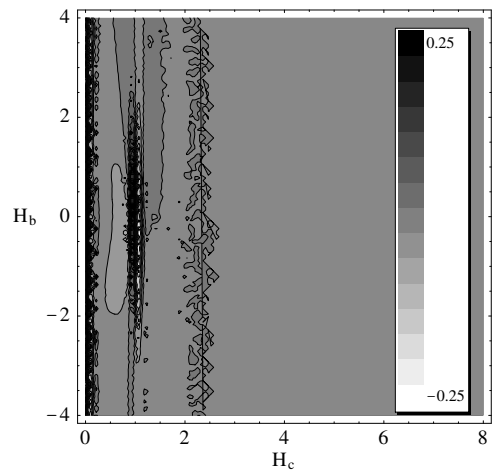}
\end{center}
\vspace*{-0.5cm}
\caption{
Absolute difference between the FORC diagram presented
in Fig.~\ref{fig1} and the corresponding SFH in
Fig.~\ref{fig2} for the three-dimensional RFIM with
$\sigma_R = 5.0$. The details are discussed in the
text.
\label{fig3}
}
\end{figure}

Figure \ref{fig2} shows the switching-field histogram
(SFH) for the RFIM.  One can see a close resemblance
with the corresponding FORC diagram presented in
Fig.~\ref{fig1}. In order to better compare FORC
diagram and SFH, in Fig.~\ref{fig3} we present the
absolute difference between both diagrams.

Even though the SFH and the FORC diagram of the RFIM
differ slightly (see Fig.~\ref{fig3}), the main
characteristics representing the underlying physical
properties of the model are the same (vertical ridge
representing multi-domain nucleation). It is
interesting that a zeroth-order reversal curve (the
major hysteresis loop) contains possibly all the
necessary information to reconstruct the first order
reversal curves of the system (the FORC diagram).

The differences found between the FORC diagram and the
SFH could be due to numerical error in the derivatives
of the FORCs because noise is amplified in numerical
derivatives considerably. In addition, the differences
could be attributed to either hysteron correlations or
the failure of the (simple) Preisach picture of
hysterons. The latter would imply that a
generalization of ``classical'' hysterons is required.

Figure \ref{fig2} also illustrates how the
re-parametrization of the major hysteresis loop in
terms of an SFH shows more details about the
microscopic structure of the system. The gained
information is similar to the information provided by
an FORC diagram, yet the computation of an SFH is {\em
considerably} faster than calculating an FORC diagrams
(generally $\sim 10^2$ times faster) and involves no
numerical derivatives of the data, thus reducing
numerical error.
 
\section{Frustrated Systems}
\label{frustr}

As the random-field Ising model is a random magnet
with no frustration, we also calculate the FORC
diagram and SFH for the 3D Edwards-Anderson Ising spin
glass \cite{binder:86} (EASG). Due to frustration, a
spin can flip more than twice along the full
hysteresis loop. With the current definition of the
SFH this is not taken into account and differences to
an FORC diagram are expected.

The Hamiltonian of the Edwards-Anderson Ising spin
glass is given by Eq.~(\ref{eq:hamilton}) where the
$J_{ij}$ are nearest-neighbor interactions chosen
according to a Gaussian distribution with zero mean
and standard deviation unity, and $h_i = 0$ $\forall
i$. $H$ represents the externally applied magnetic
field and periodic boundary conditions are applied.
For the simulations we use the zero-temperature
algorithm presented in Sec.~\ref{model}. Frustration
is introduced by the random signs of the interactions
$J_{ij}$.

\begin{figure}[tbp]
\begin{center}
\includegraphics[scale=0.38]{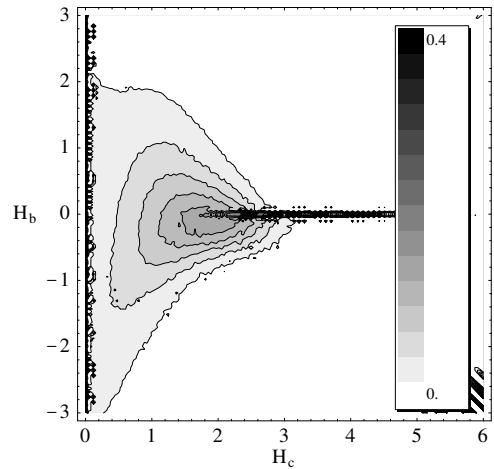}
\end{center}
\vspace*{-0.5cm}
\caption{
FORC Diagram of the EASG. Note the ridge along the
$H_c$-axis. Data for 5000 disorder realizations and $N
= 50^3$ spins.
\label{fig4}
}
\end{figure}

Figure \ref{fig4} shows the FORC diagram of the EASG.
One can see a pronounced ridge along the $H_{\rm
c}$-axis together with an asymmetric feature at small
coercivities. The underlying details of the EASG FORC
diagram have been discussed
elsewhere \cite{katzgraber:02b}.

\begin{figure}[tbp]
\begin{center}
\includegraphics[scale=0.38]{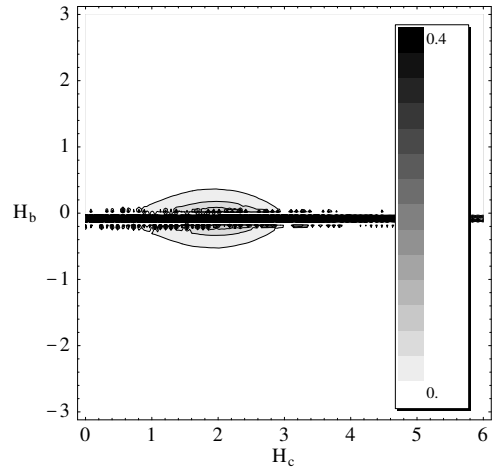}
\end{center}
\vspace*{-0.5cm}
\caption{
SFH of the EASG.  While the ridge along the $H_c$-axis
is qualitatively conserved, the SFH shows drastic
differences to the FORC diagram presented in
Fig.~\ref{fig4}. In particular, the asymmetry with
respect to the horizontal axis is lost.
\label{fig5}
}
\end{figure}

In Fig.~\ref{fig5} the SFH of the EASG is shown. Note
that the asymmetry present in the FORC diagram in
Fig.~\ref{fig4} is lost. The weight of the asymmetric
part of the FORC diagram shifts to the ridge at $H_b =
0$.

Although some of the features in the FORC diagram
(Fig.~\ref{fig4}) of the EASG are missing in the
corresponding SFH (Fig.~\ref{fig5}), the horizontal
ridge reminiscent of the underlying reversal-symmetry
of the Hamiltonian \cite{katzgraber:02b} is conserved.  
In particular, by comparing the FORC diagram and SFH
one can study the effects of frustration on the
hysteretic behavior of a spin glass.

\section{Conclusions}
\label{conclusions}

By re-parameterizing the major hysteresis loop with a
switching-field histogram we show that for systems
with no frustration (random-field Ising model) the SFH
closely resembles the FORC diagram. Small differences
can be attributed to numerical error in the
calculation of an FORC diagram, hysteron correlations,
or the breakdown of the hysteron picture.

SFHs show more details about the system than the major
hysteresis loop and are considerably faster to
computer than FORC diagrams. Therefore they are an
efficient alternative in order to study the
microscopic distributions of coercivity and bias of
the switching units.

Because the switching fields of the underlying
microscopic switching units have to be recorded for
the computation of an SFH, the method is in general 
limited to numerical studies of hysteretic systems. 
Experimental applicability might be possible with 
synthetic particulate samples \cite{liu:02} where 
the individual switching units can be traced during 
the magnetic field sweep.

We also present results on the (frustrated)
Edwards-Anderson Ising spin glass. We show that there
are clear differences between the FORC diagram and the
SFH because SFHs do not take into account multiple
switching events of the spins, a hallmark of spin
glasses. We suggest these differences can be used to
quantify the effects of frustration in FORC diagrams.

The FORC and SFH methods promise to be powerful tools to
``fingerprint'' hysteretic systems. Still, the breadth
of information they provides remains to be understood
fully.

\section*{Acknowledgments}
We would like to thank F.~P\'azm\'andi and
R.~T.~Scalettar for discussions.

\bibliography{refs}

\end{document}